\documentclass[12pt,preprint]{aastex}
\newcommand{\be}{\begin{equation}}
\newcommand{\ee}{\end{equation}}
\newcommand{\bbb}{\begin{eqnarray}}
\newcommand{\eee}{\end{eqnarray}}

\def\note #1]{{\bf #1]}}
\def\gwig{{\leavevmode\kern0.3em\raise.3ex\hbox{$>$}
\kern-0.8em\lower.7ex \hbox{$\sim$}\kern0.3em}}
\def\lwig{{\leavevmode\kern0.3em\raise.3ex\hbox{$<$}
\kern-0.8em\lower.7ex \hbox{$\sim$}\kern0.3em}}

\shorttitle{Amplitude Modulation in Time-Distance Helioseismology} 
\shortauthors{Nigam \& Kosovichev}

\begin{document}
\title{Note on Travel Time Shifts due to   Amplitude Modulation  in  Time-Distance Helioseismology Measurements}
\author{R. Nigam and A. G. Kosovichev}
\affil{W. W. Hansen  Experimental Physics Laboratory, Stanford 
University, Stanford, USA} \email{{rakesh,sasha}@quake.stanford.edu}

\begin{abstract}

Correct interpretation of acoustic travel times measured by 
time-distance helioseismology is essential to get an accurate 
understanding of the solar properties that are inferred from them. 
It has long been observed that sunspots suppress $p$-mode amplitude, 
but its implications on travel times has not been fully investigated 
so far. It has been found in test measurements using a  'masking' 
procedure, in which the solar Doppler signal in a localized quiet 
region of the Sun is artificially suppressed by a spatial function, 
and using numerical simulations that the amplitude modulations in 
combination with the phase-speed filtering may cause systematic 
shifts of acoustic travel times. To understand the properties of 
this procedure, we derive an analytical expression for the 
cross-covariance of a signal that has been modulated locally by a 
spatial function that has azimuthal symmetry,  and then filtered by 
a phase speed filter typically used in time-distance 
helioseismology. Comparing this expression to the  Gabor wavelet 
fitting formula without this effect, we find that there is a shift 
in the travel times, that is introduced by the amplitude modulation. 
The analytical model presented in this paper can be useful also for 
interpretation of travel time measurements for non-uniform 
distribution of oscillation amplitude due to observational effects.

\end{abstract}

\section{Introduction}
\label{sec:intro}

Time-distance helioseismology \citep{duv93} is a  local 
helioseismological technique that has been used to study  meridional 
flows, flows and  sound speed perturbations beneath sunspots 
\citep[e.g.,][]{kos97, gil97, zha01}. It measures the time  for a 
wavepacket to travel between any two points on the solar surface, by 
computing a temporal cross-covariance between the  Doppler time 
series at the two points. The travel time is then inverted to infer 
various  properties that are useful to map local structures of the 
sun. These results are interesting, as  they complement the results 
obtained from global helioseismology based on inversion of normal 
mode frequencies. However, many aspects of time-distance 
helioseismology are still not fully understood. In particular, it 
has been observed that sunspots suppress the $p$-mode amplitude 
appreciably, but its consequences on travel times and the properties 
derived by inverting them has largely remained unexplored.

The commonly used procedure of measuring the phase and group travel 
times of acoustic waves is based a Gabor-wavelet fitting formula 
derived by \citet{kos97} for cross-covariance of randomly excited 
oscillation modes of the quiet Sun (represented by a spherically 
symmetric model). Initially, this procedure included only a 
broad-band frequency filtering of the solar oscillation signal, 
centered at the peak of acoustic power. Later, it was modified by 
adding a phase-speed filtering procedure in order to isolate the 
first-bounce signals (direct waves without additional reflections 
from the surface) and to improve the signa-to-noise ratio 
\citep{Duvall1997}. The phase-speed filtering is important for the 
analysis of acoustic-wave packets traveling short travel distances,
e.g. less than $1$ heliographic degree ($\sim 12$ Mm). For larger 
distances the travel times can be measured without the phase-speed 
filtering, and it is found that the use of the phase-speed filtering 
does not significantly affects the measurement results. For shorter 
distances the influence of the phase-speed filtering may be 
significant, and has to be taken account. 

In this paper we study the effects of the phase-speed filtering and 
local spatial suppression of Doppler velocity signals  from a quiet 
patch on the Sun, on the travel times. In this `masking' procedure 
suggested by \citet{raj06} to simulate this effect in sunspots the 
oscillation signal of a quiet Sun region is multiplied by an 
inverted modulation function of spatial coordinates with azimuthal 
symmetry.  This function is called a mask, and is not a function of 
time. One should note that while this procedure models a spatial 
modulation of acoustic power, it does not represent the modulation 
observed in sunspots, where the amplitude of acoustic changes due do 
several physical factors, such as reduced excitation, absorption, 
changing wave speed, and spectral line formation observing 
conditions. Recently, \citet{Chou2009} attempted to quantify these 
contributions using observational data. The variations of the 
oscillation power in sunspots have not been explained by theory or 
simulations \citep[e.g.][]{Parchevsky2007}. Nevertheless, it is 
interesting to investigate the effects of the variations on the 
helioseismic travel times by using the simple `masking' model 
bearing in mind that it may not represent the the real situation in 
sunspots. One advantage of the `masking' model is that it allows 
relatively simple analytical investigation.

To understand the results of amplitude suppression or enhancement we 
theoretically model the effect of masking by computing the 
cross-covariance of a  masked signal that has been filtered by an 
appropriately designed phase speed filter. Analytical expressions 
for the cross-covariance are derived in terms of the mask, the 
filter parameters, the properties of the signal and the dispersive 
nature of the solar medium, when the the mask is azimuthally 
symmetric.  This analysis will be useful to study the effect of 
masking on travel time measurements and the properties that are 
inferred from inverting the travel times. This will be especially 
valuable in artificially mimicking  how the sunspots influence the 
travel times of  $p$ modes, and also other instrumental effects that 
corrupt the observed signal.

\section{Effect of phase-speed filtering on cross-covariance and travel times} \label{sec:analysis}

Consider a Doppler signal from a quiet patch on the Sun. Travel time 
maps for this region are computed by fitting the observed 
cross-covariance by a Gabor wavelet \citep{kos97}. Now this region 
is masked by a spatial function to induce an artificial suppression 
in amplitude. One could alternatively enhance the amplitude in a 
similar manner. Maps for both mean and difference travel times are 
computed, also by fitting a Gabor wavelet. Taking the difference 
between the quiet and masked travel time maps, one sees appreciable 
shifts in the travel times around the masked region. This is 
illustrated in the paper of \citet{raj06}. It is generally observed 
that the difference travel times show appreciable shifts compared to 
the mean travel times in the masked region.

In time-distance helioseismology we deal with acoustic waves near 
the solar surface, that are observed by measuring the line of sight 
Doppler velocity signal on the solar surface that has both radial 
and horizontal components of displacement. Without loss of 
generality, the line of sight direction is taken along the X axis. 
The signal is a sum of normal modes, and at a position $\vec R= (R, 
\theta, \phi)$ on the solar surface and time $t$ is written as, $d^x 
(\vec R, t) = \sum_{n, l, m} d_{nlm}^x(\vec R, t)$ 
\citep[e.g.]{jcd02}. For each mode,
\begin{equation}
d_{nlm}^x(\vec R, t) = d_{nlm}^x(\vec R) \exp \left 
\{i(\omega_{nlm}t - \alpha_{nlm})\right \} \label{eq1}
\end{equation}
where, the spatial part is given by projecting along the X-axis. 
\small
\begin{equation}
d_{nlm}^x(\vec R) = a_{nlm} \xi_{nl}^r(R) \left \{ {Y}^{m}_l(\theta, 
\phi) \sin\theta \cos\phi + \beta_{nl}(R) \left [\frac{ \partial 
{Y}^{m}_l(\theta, \phi) } {\partial \theta} \cos\theta \cos\phi - 
\frac {\sin\phi} {\sin\theta} \frac { \partial {Y}^{m}_l(\theta, 
\phi)} {\partial \phi} \right ] \right \}\nonumber \label{eq2}
\end{equation}
\normalsize where,
\begin{equation}
\beta_{nl}(R) = {\xi_{nl}^h(R)}/{\xi_{nl}^r(R)}   \label{eq3}
\end{equation}
where, $\theta$ is the co-latitude, $\phi$ is the longitude, $R$ is 
the solar radius,  $i = \sqrt{-1}$, the mode amplitude $a_{nlm}$, 
the phase $\alpha_{nlm}$, and $\xi_{nl}^{r}(R)$ and 
$\xi_{nl}^{h}(R)$ are the radial and horizontal components of the 
eigenfunctions respectively evaluated at $R$, and are therefore 
numbers. The spherical coordinate system is specified by unit vector 
$\hat e_r$ in the radial direction $r$, and two unit vectors $\hat 
e_{\theta}$ and $\hat e_{\phi}$ in the horizontal directions 
$\theta$ and $\phi$ respectively. Each normal mode is specified by a 
3-tuple $(l, m, n)$ of integer parameters, corresponding angular 
frequency $\omega_{nlm}$, the mode amplitude $a_{nlm}$, the phase 
$\alpha_{nlm}$. The integer $l$ denotes the degree and $m$ the 
azimuthal order, $-l \le m \le l$, of the spherical harmonic 
$Y^m_l(\theta,\phi)$, which is a function of the co-latitude 
$\theta$ and longitude $\phi$. These describe the angular structure 
of the eigenfunctions. The third integer $n$ of the 3-tuple $(l, m, 
n)$ is called the radial order. For a spherically symmetric Sun, all 
modes with the same $n$ and $l$ have the same eigenfrequency 
$\omega_{nl}$, regardless of the value of $m$.

In time-distance helioseismology we measure the travel time of wave 
packets by forming a temporal cross-covariance between the 
oscillation signals at two locations separated by an angular 
distance on the solar surface. To model this we represent the solar 
oscillations on the solar surface as a linear superposition of 
normal modes, that are band-limited in angular frequency $\omega$. 
This is achieved by replacing $a_{nlm} \xi_{nl}^r(R)$ in equation 
(\ref{eq2}) by the Gaussian frequency function $G_l(\omega)$, which 
models the amplitude of the solar modes,
\begin{equation}
G_l(\omega) = b_l \exp\left(-{(\omega - \omega_o)^2}/{\delta 
\omega^2}\right) \label{eq4}
\end{equation}
where, $b_l$ is a coefficient of $l$, which is discussed in 
\citep{nig07}. This function groups modes within a certain range of 
frequencies, which is controlled by the width $\delta \omega$, about 
a central frequency $\omega_o$ in the $\omega - l$ diagram.

A phase speed filter is applied, and modes are selected from the 
$\omega - l$ diagram to construct the cross-covariance  wave packet. 
It is specified by a Gaussian centered around a  phase speed 
$V_{ph}$ and  a width $\delta V_{ph}$ as parameters, and is given by
\begin{equation}
F_p(V_p) = \exp\left(-{\left(V_p - V_{ph}\right)^2}/{\delta 
V_{ph}^2}\right) \label{eq5}
\end{equation}
where, the phase speed $V_p = {\omega}/{L}$,  $L = \sqrt{l(l+1)} = 
k_hR$, $k_h$ is the horizontal  wave number. The role of the phase 
speed filter is to select waves with a small range of phase speeds, 
the range is specified by the width $\delta V_{ph}$. All these waves 
travel approximately  the same horizontal distance on the solar 
surface, and sample  the same vertical depth in the solar interior. 
Hence, it is crucial to select $\delta V_{ph}$ appropriately so as 
to make the cross-covariance more robust, and hence  be able to 
resolve the sub-surface structures in the Sun.

Due to the band-limited nature of the oscillation amplitudes, only 
values of  $L$ which are close to $L_o = {\omega_o}/{V_p}$ 
contribute to the sum of  equation (\ref{eq1}), and hence, following 
\citet{kos97}, we Taylor expand $L$ about the central frequency 
$\omega_o$, up to the first order:
\begin{equation}
L = L(\omega) = L \left [\omega_o + (\omega - \omega_o) \right] 
\approx L(\omega_o) + \frac{dL}{d\omega}(\omega - \omega_o) 
\label{eq6}
\end{equation}
The  equation  (\ref{eq6}) can be written in terms of the group 
velocity $U_g = {d\omega}/{dL}$ and phase velocity $V_p = 
{\omega}/{L}$, evaluated at $\omega = \omega_o$, and using the fact 
$L(\omega_o) = {\omega_o}/{V_p}$, we have
\begin{equation}
L(\omega) \approx \frac{\omega}{U_g} + \left(\frac{1}{V_p} - 
\frac{1}{U_g}\right)\omega_o \label{eq7}
\end{equation}
Likewise, the phase velocity $V_p(L, \omega)$ can be expanded about 
the point $(L_o,\omega_o)$ in the $\omega - l$ diagram to yield
\begin{equation}
F_p(L,\omega) \approx \exp\left(-{V_{ph}^2\left(L - 
\frac{\omega}{V_{ph}}\right)^2}/{\delta_f^2}\right) \label{eq8}
\end{equation}
where, $\delta_f = {\omega_o\delta V_{ph}}/{V_{ph}}$, and the filter 
width $\delta V_{ph}$ is evaluated at $(L_o, \omega_o)$, and is a 
constant.

The Taylor expansion is valid when the second order effects are 
small. These may not be small for small distances $\Delta$, when the 
waves spend most of the time in the outer layers of the Sun. In 
these layers all the solar properties change abruptly and there are 
large gradients present, so higher order terms in the Taylor 
expansion  should be retained. This could make the analytical 
calculation formidable.

The cross-covariance $\psi_{f_p}^{d^x}(\vec R_1,\vec R_2,\tau)$ for 
the  phase speed filtered Doppler signal\\ $d_{f_p}^x (R, \theta, 
\phi, t) = \sum_{n,l,m} F_p(L, \omega_{nl}) d_{nlm}^x(\vec R, t)$ as 
a function of the time lag $\tau$ is defined as
\begin{equation}
\psi_{f_p}^{d^x}(\vec R_1,\vec R_2,\tau) = \frac{1}{T} 
\int\limits_{0}^{T} d_{f_p}^x(\vec R_1, t){\bar d_{f_p}^x}(\vec R_2, 
t + \tau) dt \label{eq11}
\end{equation}
and involves integrating the  product of the projected line of sight 
filtered Doppler signals  at the two locations $\vec R_1 = (R, 
\theta_1,\phi_1)$ and  $\vec R_2 = (R, \theta_2,\phi_2)$ on the 
solar surface over a time interval $T$, that is related to the 
period of the time series being cross correlated. Here we have 
replaced $a_{nlm} \xi_{nl}^r(R)$ by the Gaussian frequency envelope 
function  $G_l(\omega)$ in  equation (\ref{eq4}).

The cross-covariance from equation (\ref{eq11}) is therefore,
\begin{equation}
\psi_{f_p}^{d^x}(\vec R_1, \vec R_2, \tau) \approx \sum_{V_p} 
\frac{2 \pi C_l}{L\sqrt{\pi\Delta}}  \frac{1}{2\pi} 
\sqrt{\frac{L}{2}} \sum_{\omega_{nl}} F_p^2(L, \omega_{nl}) 
G_l^2(\omega_{nl})[f_{+} (\omega_{nl} \tau) + f_{-} (\omega_{nl} 
\tau)] \label{eq12}
\end{equation}
where, $f_{+} (\omega \tau) = \cos(\omega \tau - L\Delta + 
\frac{\pi}{4} - \zeta)$ corresponds to the positive time lag and 
$f_{-} (\omega \tau) = \cos\left(\omega \tau + L\Delta  - 
\frac{\pi}{4} + \zeta \right)$, corresponds to the negative time 
lag. Since $\cos$ is an even function, $f_{-} (\omega \tau) = f_{+} 
(-\omega \tau)$. This approximation is valid for large $l$, small 
$\Delta$, such that $L\Delta$ is large. The phase term $\zeta$ and 
the factor $C_l$ are due to the horizontal component of the 
displacement, and depend on the location of the points $\vec R_1$ 
and $\vec R_2$, and are discussed in \citep{nig07}.

The inner sum over $\omega_{nl}$ is replaced by an an integral over 
$\omega$, and we drop the negative lag term by extending $\omega$ to 
negative values to get 
\begin{equation}
\psi_{f_p}^{d^x}(\vec R_1, \vec R_2,\tau, V_p) =  
\int\limits_{-\infty}^{\infty} d\omega \exp\left(-{2V_{ph}^2\left(L 
- \frac{\omega}{V_{ph}}\right)^2}/{\delta_f^2}\right) \exp\left 
[-\frac{2}{\delta \omega^2} (\omega - \omega_o)^2 \right ] 
\cos(\omega\tau - L\Delta - \zeta +\frac{\pi}{4}) \label{eq13}
\end{equation}
Evaluating the integral  \citep{gra94}, we get
\begin{equation}
\psi_{f_p}^{d^x}(\vec R_1, \vec R_2, \tau, V_p) = A_{f_p}(\delta 
\omega, \delta_f,\tau,\tau_g,\tau_p) \cos\left \{\omega_{f_p}(\tau - 
\tau_{f_{ph}}) +\frac{\pi}{4} \right \} \label{eq14}
\end{equation}
The shifted phase travel  time $\tau_{f_{ph}}$ due to the phase 
speed filter and horizontal component is therefore,
\begin{equation}
\tau_{f_{ph}} = \tau_p - \frac{R_{gp\varepsilon}}{1 - 
R_{gp\varepsilon}}(R_g - R_p)\tau_{ph}  + \frac{\zeta}{\omega_{f_p}} 
\label{eq15}
\end{equation}
and the shifted frequency, $\omega_{f_p} = \omega_o(1 - 
R_{gp\varepsilon})$. The amplitude scaling term is
\begin{equation}
A_{f_p} =  \sqrt {\frac{\pi}{2}} \frac{\delta \omega ~\varepsilon 
}{\sqrt {R_g^2 + \varepsilon^2}} \exp\left \{-\frac{\delta \omega^2 
\varepsilon^2}{8(R_g^2 + \varepsilon^2)}\left((\tau - \tau_g)^2
 + \frac{16\omega_o^2 R_p^2 }{\delta \omega^4 \varepsilon^2}\right)\right \}
\label{eq16}
\end{equation}
Summing equation (\ref{eq14})  over phase velocities we get the 
final cross-covariance.
\begin{equation}
\psi_{f_p}^{d^x}(\vec R_1, \vec R_2, \tau) = \sum_{V_p} \frac{2 \pi 
C_l}{L\sqrt{\pi\Delta}} \psi_{f_p}^{d^x}(\vec R_1, \vec R_2, \tau, 
V_p) \label{eq17}
\end{equation}
The dimensionless quantities $R_g = {\tau_g - 
\tau_{ph}}/{\tau_{ph}}$ and  $R_p = {\tau_p - 
\tau_{ph}}/{\tau_{ph}}$ represent the relative deviation of the 
group travel time $\tau_g = {\Delta}/{U_g}$ and phase travel time 
$\tau_p = {\Delta}/{V_p}$ respectively from the filter phase travel 
time, $\tau_{ph} =  {\Delta}/{V_{ph}}$. The filter widths appear in 
a dimensionless parameter    $\varepsilon^2 = {\delta_f^2}/{\delta 
\omega^2}$. The dispersive characteristics of the solar medium and 
the filter properties are related through the dimensionless 
parameter $R_{gp\varepsilon} = {R_gR_p}/{(R_g^2 + \varepsilon^2)}$.

Equations (\ref{eq14}-\ref{eq16}) provide a generalization of the 
Gabor wavelet fitting formula of \citet{kos97} for the phase-speed 
filtering procedure. Obviously, the filter width $\delta V_{ph}$ is 
very large, so that parameter $\epsilon \rightarrow \infty$ these 
equations are reduced to the standard fitting formula.

\section{Effect of amplitude modulation on time-distance helioseismology measurements}

\subsection{Cross-covariance for solar oscillations with spatial modulation, and travel-time shifts}

In this section, we derive a formula for the time-distance 
cross-covariance function in the presence of a localized amplitude 
masking. This provides estimates of the masking effect on the 
time-distance helioseismology measurements for various wave 
properties and observational parameters, including phase-speed 
filtering which is the major factor affecting the measurements.
 We consider a
modulation function $q(\theta)$ with azimuthal symmetry so as to 
simplify the analytical derivation. It can therefore be expanded as

\begin{equation}
q(\theta) = \sum_{l} q_l P_l(\cos \theta) \approx \sum_{l} q_l 
\sqrt{\frac{2}{\pi L\theta}} \cos \left (L\theta - \frac{\pi}{4} 
\right) \label{eq30}
\end{equation}
where, the approximation is valid for large $l$, small $\theta$, 
such that $L\theta$ is large \citep{jackson-book}.  It should be 
noted that due to the azimuthal symmetry the mask function is 
independent of $\phi$, and hence on $m$ in this expansion in 
equation (\ref{eq30}). Using orthonormality of $P_l(\cos \theta)$, 
we can compute the coefficient $q_l$ as
\begin{equation}
q_l = \frac{2l + 1}{2} \int\limits_{0}^{\pi} q(\theta) P_l(\cos 
\theta) \sin \theta ~~d \theta \label{eq31}
\end{equation}
Since the masking is carried out in a localized region we define the 
mask function that we apply to the signal as
\begin{equation}
Q(\theta) = 1 + s ~ q(\theta) = \sum_{l} Q_l P_l(\cos \theta) 
\approx \sum_{l} Q_l \sqrt{\frac{2}{\pi L\theta}} \cos \left 
(L\theta - \frac{\pi}{4} \right) \label{eq32}
\end{equation}
where, $s$ scales the mask function $q(\theta)$,  and  is positive 
for enhancing the signal and negative for suppressing it. The 
coefficient $Q_l$ is
\begin{equation}
Q_l = \frac{2l + 1}{2} \int\limits_{0}^{\pi} Q(\theta) P_l(\cos 
\theta) \sin \theta ~~d \theta \label{eq33}
\end{equation}
Masking takes place in the spatial domain and is independent of 
time. It consists of multiplying the  signal $d^x (R, \theta, \phi, 
t)$ with the mask function $Q(\theta)$.

The mask function $Q(\theta)$  given in equation (\ref{eq32}),  and 
it is used to spatially modulate the signal  $d^x(\vec R, t)$ 
resulting in the masked signal $d_Q^x(\vec R, t)$,
\begin{equation}
d_Q^x(\vec R, t) = Q(\theta) d^x(\vec R, t)  =  [1 + s ~ 
q(\theta)]d^x(\vec R, t)
 = d^x(\vec R, t) + s ~ q(\theta)d^x(\vec R, t) =  d^x(\vec R, t) + s ~ d_q^x(\vec R, t)
\label{eq34}
\end{equation}
Where, $d_q^x(\vec R, t) = q(\theta) d^x(\vec R, t)$ is the 
spatially modulated signal. The effect of masking is seen when we 
phase speed filter the masked signal. In the absence of a phase 
speed filter we do not observe any masking effect in the 
time-distance cross-covariance.

Phase speed filtering of the masked signal leads to
\begin{equation}
d_{f_{pQ}}^x(\vec R, t) = \sum_{n,l,m} G_l(\omega_{nl})  F_p(L, 
\omega_{nl}) Q(\theta) d_{nlm}^x(\vec R, t) \label{eq35}
\end{equation}
The cross-covariance of the masked filtered signal is
\begin{equation}
\psi_{f_{pQ}}^{d^x}(\vec R_1,\vec R_2,\tau) =  \frac{1}{T} 
\int\limits_{0}^{T} d_{f_{pQ}}^x(\vec R_1, t){\bar 
d_{f_{pQ}}^x}(\vec R_2, t + \tau) dt \label{eq36}
\end{equation}
Substituting the expression for $d_{f_{pQ}}^x(\vec R, t)$ from 
equation (\ref{eq35}) into equation (\ref{eq36}) and using equation 
(\ref{eq32}) leads to
\begin{equation}
\psi_{f_{pQ}}^{d^x}(\vec R_1,\vec R_2,\tau) = \psi_{f_p}^{d^x}(\vec 
R_1,\vec R_2,\tau) + s~ \psi_{f_p,f_{pq}}^{d^x}(\vec R_1,\vec 
R_2,\tau) + s~ \psi_{f_{pq},f_p}^{d^x}(\vec R_1,\vec R_2,\tau) + s^2 
~ \psi_{f_{pq}}^{d^x}(\vec R_1,\vec R_2,\tau) \label{eq37}
\end{equation}
This is the final expression for the time-distance cross-covariance 
function with amplitude masking. It contains terms that result from 
the interaction of the phase speed filter and the mask function. The 
term $\psi_{f_p}^{d^x}(\vec R_1,\vec R_2,\tau)$ is due to the phase 
speed filter alone, and does not contain the effect of the mask. The 
term $\psi_{f_p,f_{pq}}^{d^x}$ is obtained by cross correlating the 
filtered signal with the modulated filtered signal, 
$\psi_{f_{pq},f_p}$ is the cross covariance of the modulated 
filtered signal with the filtered signal. The last term 
$\psi_{f_{pq}}$ is got by cross correlating the modulated filtered 
signal at both the points $\vec R_1$ and $\vec R_2$.


The cross-covariance from equation (\ref{eq11}) is therefore,
\begin{equation}
\psi_{f_p}^{d^x}(\vec R_1,\vec R_2,\tau) = \frac{1}{T} 
\int\limits_{0}^{T} d_{f_p}^x(\vec R_1, t){\bar d_{f_p}^x}(\vec R_2, 
t + \tau) dt \label{eq38}
\end{equation}
\begin{equation}
\psi_{f_p}^{d^x}(\vec R_1, \vec R_2, \tau) = \sum_{n,l} F_p^2(L, 
\omega_{nl}) G_l^2(\omega_{nl}) \cos(\omega_{nl}\tau) d_{nl}^x(\vec 
R_1, \vec R_2) \label{eq39}
\end{equation}
Where, $d_{nl}^x(\vec R_1, \vec R_2) = \sqrt{\frac{2}{\pi L\Delta}} 
\cos(L \Delta + \zeta)$ is the $m$-averaged part of the spatial 
signal in the cross covariance \citep{nig07}. We have on combining 
the cosine terms,
\begin{equation}
\psi_{f_p}^{d^x}(\vec R_1, \vec R_2, \tau) = \sum_{V_p} \frac{2 \pi 
C_l}{L \sqrt{\pi\Delta}}  \frac{1}{2\pi} \sqrt{\frac{L}{2}} 
\sum_{\omega_{nl}} F_p^2(L, \omega_{nl}) G_l^2(\omega_{nl})[f_{+} 
(\omega_{nl} \tau) + f_{-} (\omega_{nl} \tau)] \label{eq40}
\end{equation}
The phase factor $\zeta$ is due to the horizontal component of the 
displacement, and depends on the location of the points being cross 
correlated.

In  equation (\ref{eq40}) we replace the dummy variable 
$\omega_{nl}$ by $\omega$, and drop the negative time lag term by 
extending the sum over $\omega$ to negative frequencies, we get
\begin{equation}
\psi_{f_p}^{d^x}(\vec R_1, \vec R_2, \tau) = \sum_{V_p} \frac{2 \pi 
C_l}{L\sqrt{\pi\Delta}}  \frac{1}{2\pi} \sqrt{\frac{L}{2}} 
\sum_{\omega} F_p^2(L, \omega) G_l^2(\omega) \cos \left (\omega \tau 
- L\Delta + \frac{\pi}{4} - \zeta \right ) \label{eq41}
\end{equation}
The inner sum in equation (\ref{eq41}) is written as
\begin{equation}
\psi_{f_p}^{d^x}(\vec R_1, \vec R_2, \tau, V_p) = \sum_{\omega} 
F_p^2(L, \omega) G_l^2(\omega) \cos \left (\omega \tau - L\Delta + 
\frac{\pi}{4} - \zeta \right ) \label{eq42}
\end{equation}
Multiplying with the mask function the different cross covariance 
are,
\begin{equation}
\psi_{f_p,f_{pq}}^{d^x}(\vec R_1,\vec R_2,\tau) = \frac{1}{T} 
\int\limits_{0}^{T} d_{f_p}^x(\vec R_1, t){\bar d_{f_{pq}}^x}(\vec 
R_2, t + \tau) dt \label{eq43}
\end{equation}
\begin{equation}
\psi_{f_p,f_{pq}}^{d^x} (\vec R_1, \vec R_2, \tau) = \sum_{n,l} 
F_p^2(L, \omega_{nl}) G_l^2(\omega_{nl}) \cos(\omega_{nl}\tau) 
q(\theta_2) d_{nl}^x(\vec R_1, \vec R_2), \label{eq44}
\end{equation}
where $\theta_2$ is colatitude of $\vec R_2$.

 Substituting the expansion for $q(\theta_2) = \sum_{l} q_l 
P_l(\cos \theta_2)$ into equation (\ref{eq44}) we get
\begin{equation}
\psi_{f_p,f_{pq}}^{d^x} (\vec R_1, \vec R_2, \tau) = \sum_{n,l} q_l 
F_p^2(L, \omega_{nl}) G_l^2(\omega_{nl}) \cos(\omega_{nl}\tau) 
P_l(\cos \theta_2) d_{nl}^x(\vec R_1, \vec R_2) \label{eq45}
\end{equation}

Substituting the asymptotic expansion for $P_l(\cos \theta_2) 
\approx \sqrt{\frac{2}{\pi L\theta_2}} \cos(L\theta_2 - 
\frac{\pi}{4})$ into equation (\ref{eq45}) and combining the cosine 
terms we get
\begin{equation}
\psi_{f_p,f_{pq}}^{d^x} (\vec R_1, \vec R_2, \tau) = \sum_{V_p} 
\frac{2 \pi C_l}{\sqrt{L\pi\Delta}}  \frac{1}{2\pi} 
\sqrt{\frac{L}{2}} \sqrt{\frac{2}{\pi L\theta_2}} q_l 
\sum_{\omega_{nl}} F_p^2(L, \omega_{nl}) G_l^2(\omega_{nl})  
\cos(L\theta_2 - \frac{\pi}{4}) [f_{+} (\omega_{nl} \tau) + f_{-} 
(\omega_{nl} \tau)]
\nonumber \label{eq46}
\end{equation}
\normalsize

Combining the cosine terms in equation (\ref{eq46}) and letting 
$f_{+d2}(\omega \tau)  = \cos(\omega \tau - L\Delta_{2-} - \zeta)$, 
$f_{-d2}(\omega \tau)  = \cos(\omega \tau + L\Delta_{2-} + \zeta) = 
f_{+d2}(-\omega \tau)$, $f_{+p2}(\omega \tau)  = \cos(\omega \tau - 
L\Delta_{2+} - \zeta + \frac{\pi}{2})$, $f_{-p2}(\omega \tau) = 
f_{+p2}(-\omega \tau)$, $\Delta_{2+} = \Delta + \theta_2$, 
$\Delta_{2-} = \Delta - \theta_2$, we get
\begin{equation}
\psi_{f_p,f_{pq}}^{d^x} (\vec R_1,\vec R_2,\tau, V_p) = 
\sum_{\omega_{nl}} F_p^2(L, \omega_{nl}) G_l^2(\omega_{nl}) (f_{+d2} 
+ f_{-d2} + f_{+p2} + f_{-p2})/2 \label{eq47}
\end{equation}
In equation (\ref{eq47}) the functions $f$ are evaluated at 
$\omega_{nl}$. The sum can be converted into an integral  over 
$\omega$ as before after dropping the negative lag terms $f_{-d}$ 
and $f_{-p}$, and extending $\omega$ to take negative values we get 
a sum of two Gabor wavelets. This shifts, the various travel times 
to $\tau_{p2+} = {\Delta_{2+}}/{V_p} = \tau_p + \tau_{pq2}$, 
$\tau_{pq2} = {\theta_2}/{V_p}$, $\tau_{g2} = {\Delta_{2+}}/{U_g} = 
\tau_g + \tau_{gq2}$, $\tau_{gq2}  = {\theta_2}/{U_g}$ and 
$\tau_{ph2} = {\Delta_{2+}}/{V_{ph}} = \tau_{ph} + \tau_{phq2}$, 
$\tau_{phq2} = {\theta_2}/{V_{ph}}$. These shifts in travel times 
are related to the mask position. Hence, $R_g$, $R_p$ and 
$R_{gp\varepsilon}$ change to $R_{g2+}$, $R_{p2+}$ and $R_{gp 
\varepsilon 2+}$ respectively, with the usual definitions. Similarly 
we have travel times for $\Delta_{2-}$: $\tau_{p2-} = \tau_p - 
\tau_{pq2}$, $\tau_{g2-} = \tau_g - \tau_{gq2}$, $\tau_{ph2-} = 
\tau_{ph} - \tau_{phq2}$. Therefore,
\begin{equation}
\psi_{f_p,f_{pq}}^{d^x} (\vec R_1, \vec R_2, \tau) = \sum_{V_p} 
\frac{2 \pi C_l}{\sqrt{L\pi\Delta}}   \sqrt{\frac{2}{\pi L\theta_2}} 
q_l \psi_{f_p,f_{pq}} (\vec R_1,\vec R_2,\tau, V_p) \label{eq48}
\end{equation}
\begin{eqnarray}
 2 \psi_{f_p,f_{pq}} (\vec R_1,\vec R_2,\tau, V_p) = & A_{f_p}(\tau_{g2-},\tau_{p2-})\cos \left \{\omega_{f_{p2-}}(\tau - \tau_{f_{ph2-}})\right \}
+ \\ & A_{f_p}(\tau_{g2+},\tau_{p2+})\cos \left 
\{\omega_{f_{p2+}}(\tau - \tau_{f_{ph2+}}) + \frac{\pi}{2} \right 
\}\nonumber \label{eq48a}
\end{eqnarray}
 Similarly  for the other terms
\begin{equation}
\psi_{f_{pq},f_p}^{d^x} (\vec R_1,\vec R_2,\tau) = \frac{1}{T} 
\int\limits_{0}^{T} d_{f_{pq}}^x(\vec R_1, t){\bar d_{f_p}^x}(\vec 
R_2, t + \tau) dt \label{eq49}
\end{equation}
\begin{equation}
\psi_{f_{pq},f_p}^{d^x} (\vec R_1,\vec R_2,\tau) = \sum_{V_p}  
\frac{2 \pi C_l}{\sqrt{L\pi\Delta}}  \frac{1}{2\pi} 
\sqrt{\frac{L}{2}}  q_l \sum_{\omega_{nl}} F_p^2(L, \omega_{nl}) 
G_l^2(\omega_{nl}) P_l(\cos \theta_1)[f_{+} (\omega_{nl} \tau) + 
f_{-} (\omega_{nl} \tau)]
\label{eq50}
\end{equation}
Simplifying in a similar manner the inner sum  of  equation 
(\ref{eq50}) is, 
\begin{eqnarray}
2 \psi_{f_{pq},f_p}^{d^x} (\vec R_1,\vec R_2,\tau, V_p) = 
A_{f_p}(\tau_{g1-},\tau_{p1-})\cos \left \{\omega_{f_{p1-}}(\tau - 
\tau_{f_{ph1-}})\right \} + \\
A_{f_p}(\tau_{g1+},\tau_{p1+})\cos \left \{\omega_{f_{p1+}}(\tau - 
\tau_{f_{ph1+}}) + \frac{\pi}{2} \right \},\nonumber \label{eq51}
\end{eqnarray}
where subscript 1 in the travel times corresponds to the mask 
position $\theta_1$ of $\vec R_1$. Therefore,
\begin{equation}
\psi_{f_{pq},f_p}^{d^x} (\vec R_1,\vec R_2,\tau) = \sum_{V_p} 
\frac{2 \pi C_l}{L\sqrt{\pi\Delta}} \sqrt{\frac{2}{\pi L\theta_1}} 
q_l \psi_{f_{pq},f_p} (\vec R_1,\vec R_2,\tau, V_p) \label{eq52}
\end{equation}
Finally,
\begin{equation}
\psi_{f_{pq}}^{d^x} (\vec R_1,\vec R_2,\tau) = \frac{1}{T} 
\int\limits_{0}^{T} d_{f_{pq}}^x(\vec R_1, t){\bar 
d_{f_{pq}}^x}(\vec R_2, t + \tau) dt \label{eq53}
\end{equation}
\begin{equation}
\psi_{f_{pq}}^{d^x} (\vec R_1,\vec R_2,\tau) = \sum_{V_p} \frac{2 
\pi C_l}{L\sqrt{\pi\Delta}}  \frac{1}{2\pi} \sqrt{\frac{L}{2}} q_l^2 
\sum_{\omega_{nl}} F_p^2(L, \omega_{nl}) G_l^2(\omega_{nl}) P_l(\cos 
\theta_1) P_l(\cos \theta_2) [f_{+} (\omega_{nl} \tau) + f_{-} 
(\omega_{nl} \tau)]\nonumber \label{eq54}
\end{equation}
 Combining the various cosine terms in the inner sum of 
equation (\ref{eq54}) we obtain,
\begin{equation}
\psi_{f_{pq}}^{d^x} (\vec R_1,\vec R_2,\tau, V_p) = \sum_{\omega} 
F_p^2(L, \omega) G_l^2(\omega) [f_{+1} + f_{-1} + f_{+2} + f_{-2} + 
f_{+3} + f_{-3} + f_{+4} + f_{-4}]/4, \label{eq55}
\end{equation}
where $f_{+1} (\omega  \tau) = \cos(\omega \tau - L\Delta_{+12-}  - 
\frac{\pi}{4} - \zeta)$, $f_{-1} (\omega  \tau) = f_{+1} (-\omega  
\tau)$, $f_{+2} (\omega  \tau) = \cos(\omega \tau - L\Delta_{+12+} +  
\frac{3\pi}{4} - \zeta)$, $f_{-2} (\omega  \tau) = f_{+2} (-\omega  
\tau)$, $f_{+3} (\omega  \tau) = \cos(\omega \tau - L\Delta_{-12-} +  
\frac{\pi}{4} - \zeta)$, $f_{-3} (\omega  \tau) = f_{+3} (-\omega  
\tau)$, $f_{+4} (\omega  \tau) = \cos(\omega \tau - L\Delta_{-12+} +  
\frac{\pi}{4} - \zeta)$, $f_{-4} (\omega  \tau) = f_{+4} (-\omega  
\tau)$, $\Delta_{+12+} = \Delta + (\theta_1 + \theta_2)$, 
$\Delta_{+12-} = \Delta - (\theta_1 + \theta_2)$, $\Delta_{-12-} = 
\Delta - (\theta_1 - \theta_2)$, $\Delta_{-12+} = \Delta + (\theta_1 
- \theta_2)$. The sum can be converted into an integral as before, 
and the resulting expression is a sum of four Gabor wavelets given 
by
\begin{equation}
g_1 (\vec R_1,\vec R_2,\tau, V_p) = 
A_{f_p}(\tau_{g+12-},\tau_{p+12-})\cos \left 
\{\omega_{f_{p+12-}}(\tau - \tau_{f_{ph+12-}})\right \} 
\label{eq51a}
\end{equation}
\begin{equation}
g_2 (\vec R_1,\vec R_2,\tau, V_p) = 
A_{f_p}(\tau_{g+12+},\tau_{p+12+})\cos \left 
\{\omega_{f_{p+12+}}(\tau - \tau_{f_{ph+12+}})\right \} 
\label{eq51b}
\end{equation}
\begin{equation}
g_3 (\vec R_1,\vec R_2,\tau, V_p) = 
A_{f_p}(\tau_{g-12-},\tau_{p-12-})\cos \left 
\{\omega_{f_{p-12-}}(\tau - \tau_{f_{ph-12-}})\right \} 
\label{eq51c}
\end{equation}
\begin{equation}
g_4(\vec R_1,\vec R_2,\tau, V_p) = 
A_{f_p}(\tau_{g-12+},\tau_{p-12+})\cos \left 
\{\omega_{f_{p-12+}}(\tau - \tau_{f_{ph-12+}})\right \} 
\label{eq51d}
\end{equation}
This shifts, the various travel times to $\tau_{+p12-} = 
{\Delta_{+12-}}/{V_p} = \tau_p - \tau_{pq1} - \tau_{pq2}, 
\tau_{+g12-} = {\Delta_{+12-}}/{U_g} = \tau_g - \tau_{gq1} - 
\tau_{gq2}$ and $\tau_{+ph12-} = {\Delta_{+12-}}/{V_{ph}} = 
\tau_{ph} - \tau_{phq1} - \tau_{phq2}$. These shifts in travel times 
are related to the mask position. Hence, $R_g, R_p$ and 
$R_{gp\varepsilon}$ change to $R_{+g12-}, R_{+p12-}$ and 
$R_{+gp\varepsilon12-}$ respectively, with the usual definitions. 
Similarly, the other combinations can be defined.

Therefore,
\begin{equation}
\psi_{f_{pq}}^{d^x} (\vec R_1,\vec R_2,\tau) = \sum_{V_p} \frac{2 
\pi C_l}{L\sqrt{\pi\Delta}} \frac{2}{\pi L \sqrt{\theta_1 \theta_2}} 
q_l^2 \psi_{f_{pq}} (\vec R_1,\vec R_2,\tau, V_p) \label{eq56}
\end{equation}
The cross covariance $\psi_{f_{pq}} (\vec R_1,\vec R_2,\tau, V_p)$ 
is the sum of the four Gabor wavelets that depend on the masking 
parameters and is given by
\begin{equation}
4 \psi_{f_{pq}}^{d^x} (\vec R_1,\vec R_2,\tau, V_p) = g_1 (\vec 
R_1,\vec R_2,\tau, V_p) + g_2 (\vec R_1,\vec R_2,\tau, V_p) + g_3 
(\vec R_1,\vec R_2,\tau, V_p) + g_4 (\vec R_1,\vec R_2,\tau, V_p) 
\label{eq57}
\end{equation}
Similarly, the cross covariance for the masking function $Q(\theta)$ 
can be found with $q_l$ repalced by $Q_l$, and is given by
\begin{equation}
\psi_{f_{pQ}}^{d^x} (\vec R_1,\vec R_2,\tau)  = \sum_{V_p} \frac{2 
\pi C_l}{L\sqrt{\pi\Delta}} \frac{2}{\pi L \sqrt{\theta_1 \theta_2}} 
Q_l^2 \psi_{f_{pQ}} (\vec R_1,\vec R_2,\tau, V_p), \label{eq58}
\end{equation}
where the coefficient $Q_l$ can be calculated from equation  
(\ref{eq33}) for the mask function $Q(\theta)$. These equations  
which are nothing but the sum of Gabor wavelets can be used for the 
fitting. Comparing the different equations we find that the form of 
the Gabor wavelet is retained  when a mask function with azimuthal 
symmetry is used. However, masking introduces shifts in the group 
and phase travel time by modifying the angular distance $\Delta$.

\begin{figure}[!ht]
\epsscale{0.8} \plotone{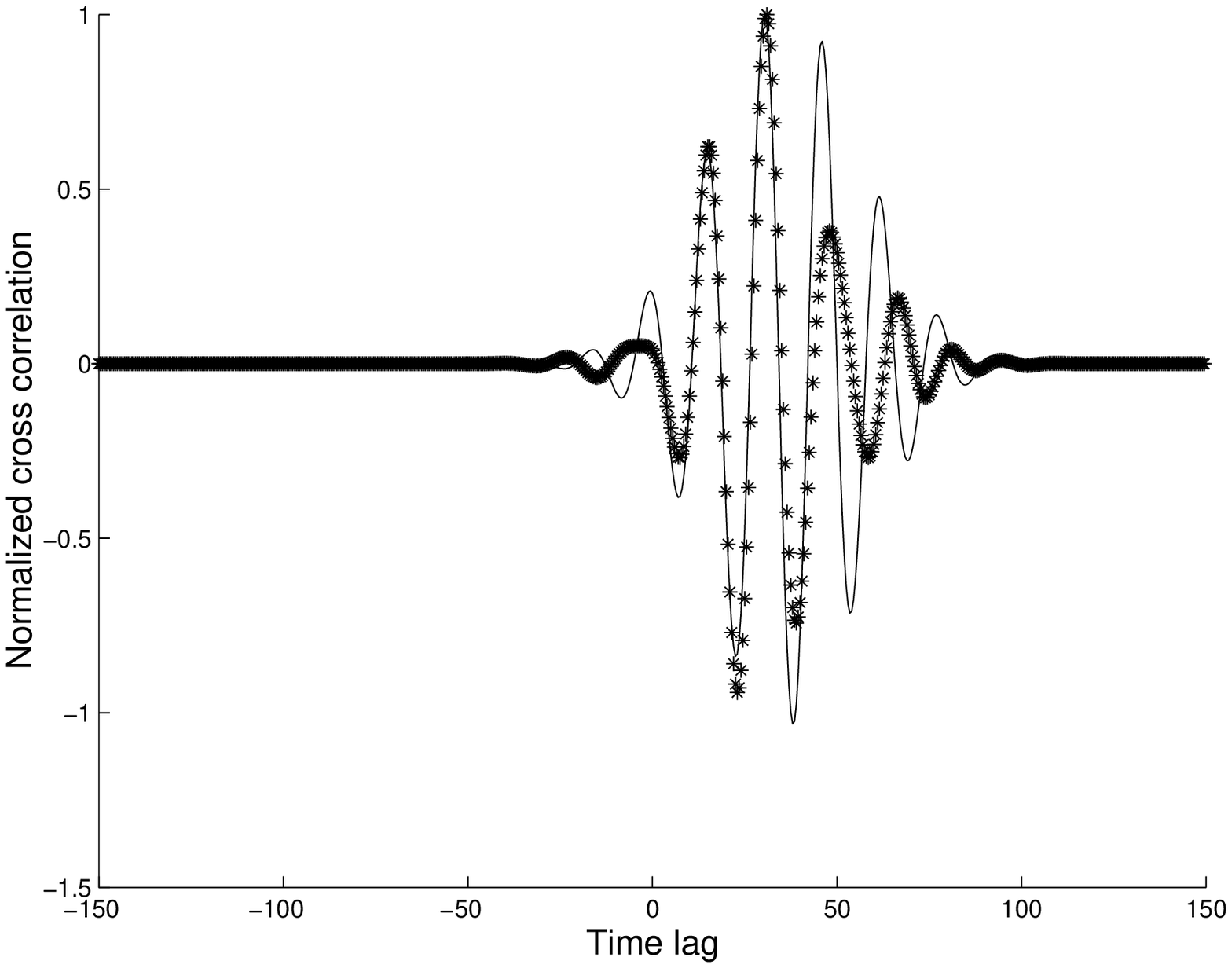} \caption{An example of the 
cross-covariance function normalized by its maximum value as a 
function of the time lag. The solid line represents no masking that 
is computed from  equation (\ref{eq14}),
 while the star (*) line represents the effect of masking, calculated from equation (\ref{eq57}). It can been seen that there
is a shift in the travel times as a result of masking.} \label{fg1}
\end{figure}

The formula  in equation (\ref{eq57}) is the masked 
cross-covariance. There are few extra parameters that represent the 
shifts in the phase and group travel times, due to the masking 
process that depend on the angular positions of the mask function. 
The coefficients $q_l$ and $Q_l$ represent the functional form of 
the mask functions. Similar dependence is seen in \citep{raj06}. A 
detailed numerical investigation of the amplitude modulation effects 
is outside the scope of this paper. In Figure~\ref{fg1}, we just 
give an example of the expression in equation (\ref{eq57}) plotted 
for a particular mask position and compared with equation 
(\ref{eq14}) without masking. We observe shifts in the cross 
covariance and hence the travel times change due to the masking 
process. To investigate to effect of the shape of the mask function, 
different values of $Q_l$ need to be included when computing the sum 
in equation (\ref{eq58}). In this derivation we assumed that the 
modulation function exhibits azimuthal symmetry. For a general  
modulation function, the analytical approach becomes difficult, and 
numerical methods must be employed.

In the model presented here, we assumed that the solar p modes have 
a narrow Gaussian amplitude function, that is peaked at $\omega_o$. 
Using this fact we evaluate the phase shift factor due to the 
horizontal component at this frequency \citep{nig07}. Hence to a 
first order approximation, the phase shift is not affected by the 
masking procedure. In order to see the effect of masking on the 
horizontal component, we have to retain the frequency dependence of 
the  phase shift factor when evaluating the integral in $\omega$. 
This will make the analytical approach intractable and the integral 
will have to be evaluated numerically.

\subsection{Phase-speed filtering and amplitude modulation (masking) do not commute}

In the previous section the signal $d^x(\vec R, t)$ was first masked 
by $Q(\theta)$ and then phase speed filtered by $F_p(L, 
\omega_{nl})$. The signal is,
\begin{equation}
d_{f_{pQ}}^x(\vec R, t) = \sum_{n,l,m} G_l(\omega_{nl})  F_p(L, 
\omega_{nl}) Q(\theta) d_{nlm}^x(\vec R, t) \label{eq59}
\end{equation}
Expanding $Q(\theta)$ from equation (\ref{eq32})  leads to
\begin{equation}
d_{f_{pQ}}^x(\vec R, t) = \sum_{n,l,m} Q_l G_l(\omega_{nl})  F_p(L, 
\omega_{nl}) P_l(\cos \theta) d_{nlm}^x(\vec R, t) \label{eq60}
\end{equation}
We immediately see that this corrupts the eigenfunction due to the 
$P_l(\cos \theta)$  term in the sum in equation (\ref{eq60}) . The 
resulting cross-covariance was $\psi_{f_{pQ}}^{d^x} (\vec R_1,\vec 
R_2,\tau)$. We now reverse the order, first phase speed filter the 
signal $d^x (\vec R, t)$ to get $d_{f_p}^x (\vec R, t)$ and then 
mask to get $d_{f_{Qp}}^x(\vec R_1,\vec R_2,\tau) = 
Q(\theta)d_{f_p}^x (\vec R,t)$. Computing the cross-covariance 
$\psi_{f_{Qp}}^{d^x} (\vec R_1,\vec R_2,\tau)$ for $d_{f_{Qp}}^{d^x} 
(\vec R,t)$,
\begin{equation}
\psi_{f_{Qp}}^{d^x}(\vec R_1,\vec R_2,\tau) = \frac{1}{T} 
\int\limits_{0}^{T} d_{f_{Qp}}^x (\vec R_1, t) \bar 
{d}_{f_{Qp}}^x(\vec R_2, t + \tau) dt \label{eq61}
\end{equation}
This results in $\psi_{f_{Qp}}^{d^x}(\vec R_1,\vec R_2,\tau) = 
Q(\theta_1)Q(\theta_2) \psi_{f_p}^{d^x}(\vec R_1,\vec R_2,\tau)$, 
which is different from the expression for $\psi_{f_{pQ}}^{d^x}(\vec 
R_1,\vec R_2,\tau)$ in equation (\ref{eq37}). Hence the order of 
masking and phase speed filtering are not interchangeable. Also we 
observe that phase speed filtering followed by masking results in 
the cross-covariance being just scaled in amplitude by the product 
of the mask functions $ Q(\theta_1)Q(\theta_2)$, the travel times 
being unaffected by the process of masking. This is because masking 
changes the mode eigenfunctions due to the presence of an additional 
term $P_l(\cos \theta)$ in equation (\ref{eq60}).  Of course, the 
modified eigenfunctions no longer satisfy the original wave equation 
for solar oscillations. Thus, this procedure cannot be considered as 
a physical model of the amplitude variations observed in sunspots. 
Our calculations show that the phase-speed filtering procedure of 
the masked signal leads to spurious shifts in travel times due to 
mixing of different $l$ modes by the phase-speed filter. These 
shifts can occur if the masking is attributed to instrumental 
effects. However, this does not mean that the amplitude modulation 
due to the physical effects in sunspots results in similar effects 
in the travel-time measurements. This must be studied using 
realistic MHD models of sunspots.

\subsection{Effect of Gaussian frequency filtering}

Masking is observed only when the data is phase speed filtered after 
multiplying the signal with the mask function. In this section we 
show that just using a  Gaussian frequency filter without a phase 
speed filter leads to no masking. Consider a   masked signal 
$d_Q^x(\vec R, t)$ that is  filtered by a Gaussian frequency filter 
$G(\omega) = \exp\left(-{(\omega - \omega_o)^2}/{\delta 
\omega^2}\right)$. The mask function $Q(\theta)$ is independent of 
$l$, and  since the Gaussian frequency filter $G(\omega)$  does not 
depend on $l$, unlike the phase speed filter, the mask function 
$Q(\theta)$ can be pulled out of the sum  in equation (\ref{eq62})

\begin{equation}
d_{f_{gQ}}^x(\vec R, t) = \sum_{n,l,m} G(\omega_{nl}) Q(\theta) 
d_{nlm}^x(\vec R, t) = Q(\theta) d_{fg}^x (\vec R, t) \label{eq62}
\end{equation}
where, $d_{fg}^x (\vec R, t) = \sum_{n,l,m} G(\omega_{nl}) 
d_{nlm}^x(\vec R, t)$ is the signal filtered by a Gaussian frequency 
filter. We see that the frequency filtered signal $d_{fg}^x (\vec R, 
t)$ is scaled by the mask function $Q(\theta)$, hence the 
cross-covariance $\psi_{f_{gQ}}^{d^x}(\vec R_1,\vec R_2,\tau)$ of 
$d_{f_{gQ}}^x(\vec R, t)$ is
\begin{equation}
\psi_{f_{gQ}}^{d^x}(\vec R_1,\vec R_2,\tau) = Q(\theta_1) 
Q(\theta_2) \psi_{f_{g}}^{d^x}(\vec R_1,\vec R_2,\tau) \label{eq63}
\end{equation}
where, $\psi_{f_{g}}^{d^x}(\vec R_1,\vec R_2,\tau)$ is the 
cross-covariance of the Gaussian filtered signal $d_{fg}^x (\vec R, 
t)$. We see that the effect of masking on the travel times is not 
observed in $\psi_{f_{gQ}}^{d^x}(\vec R_1,\vec R_2,\tau)$. It only 
undergoes a scaling of its amplitude by $Q(\theta_1) Q(\theta_2)$. 
So the effect of masking is not observed when the phase speed filter 
is absent, and we apply only a Gaussian frequency filter to the 
data.

\section{Conclusion}

In this paper we give an explanation as to why the localized spatial 
suppression or  enhancement (masking) of the acoustic signal 
followed by  phase speed filtering can appreciably shift the 
measured travel times, and cause systematic errors in time-distance 
inversions. Using only a  frequency filter or reversing the 
operation of suppression (or enhancement) and phase speed filtering 
does not shift the travel times, but merely scales the cross 
covariance. Hence, the operations of masking and phase speed 
filtering are non-commutative. To explain this we develop a  model 
and derive a new analytical formula for the cross-covariance in 
terms of the masking parameters, when the mask function has 
azimuthal symmetry. The reason for this is due to the fact that 
masking changes the spatial mode eigenfunctions, which when phase 
speed filtered, leads to a mixing of spatial modes, and is 
responsible for the travel time shifts. This may be useful to mimic 
amplitude changes in sunspots due to pure observational reasons, 
such as caused by altering the height of formation of the spectral 
line used for helioseismology measurements, and also to correct for 
the shifted travel times due to such effects. However, contrary to 
\citet{raj06} suggestion, the procedure of masking oscillations of 
the quite Sun cannot model the amplitude reduction in sunspot due to 
physical mechanisms (e.g. changes in emissivity or wave absorption), 
and thus their recommendations of correcting the amplitude reduction 
by reversed masking may cause artificial shifts in observed travel 
times, and thus must be taken with caution.

\acknowledgments This research is supported by the NASA SOHO/MDI 
grant to Stanford University.

\clearpage

\begin{thebibliography}{}



\bibitem[Chou et al.(2009)]{Chou2009} Chou, D.-Y., Liang, Z.-C., 
Yang, M.-H., Zhao, H., \& Sun, M.-T.\ 2009, \apjl, 696, L106 




\bibitem[Christensen-Dalsgaard(2002)]{jcd02} Christensen-Dalsgaard, J. 2002, {\it Rev. Mod. Phys.}, 74, 1073


\bibitem[Duvall et al.(1993)]{duv93} Duvall,  T. L., Jr., Jeffries, S. M.,
Harvey, J. W., and Pomerantz, M. A. 1993, \nat, 362, 430

\bibitem[Duvall et al.(1997)]{Duvall1997} Duvall, T.~L., Jr., et 
al.\ 1997, \solphys, 170, 63 


\bibitem[Giles et al.(1997)]{gil97} Giles, P. M., Duvall, T. L., Jr.,
Scherrer, P. H., \& Bogart, R. S. 1997, \nat, 390, 52


\bibitem[Gradshteyn \& Ryzhik(1994)]{gra94} Gradshteyn, I. S., \& Ryzhik,
I. M. 1994, page 531, Table of Integrals, Series, and Products, 
Academic Press, San Diego, fifth edition

\bibitem[Jackson(1999)]{jackson-book} Jackson, J. D., 1999, Classical Electrodynamics, 3rd edition
(New York: John Wiley \& Sons)


\bibitem[Kosovichev \& Duvall(1997)]{kos97} Kosovichev, A. G., \&  Duvall,
T. L. Jr., 1997, in F. Pijpers, J. Christensen Dalsgaard and C. S. 
Rosenthal (eds.), Solar Convection and Oscillations and Their 
Relationship, Proceedings of SCORe96 Workshop, pp 241-260, Kluwer, 
Dordrecht.

\bibitem[Nigam et al.(2007)]{nig07} Nigam, R., Kosovichev, 
A.~G., \& Scherrer, P.~H.\ 2007, \apj, 659, 1736 



\bibitem[Parchevsky 
\& Kosovichev(2007)]{Parchevsky2007} Parchevsky, K.~V., \& 
Kosovichev, A.~G.\ 2007, \apjl, 666, L53 


\bibitem[Rajaguru et al.(2006)]{raj06} Rajaguru, S. P., Birch, A. C., Duvall, T. L., Jr., Thompson, M. J.,  \& Zhao, J. 2006, \apj, 646, 543



\bibitem[Zhao, Kosovichev, \& Duvall(2001)]{zha01} Zhao, J., Kosovichev,
A. G., \& Duvall, T. L., Jr. 2001, \apj, 557, 384

\end{thebibliography}
\end{document}